\documentclass[%
 pra,
 superscriptaddress,
 amsmath,amssymb,
reprint,%
longbibliography,
]{revtex4-1}

\usepackage{hyperref}
\usepackage{amssymb}
\usepackage{aeguill}
\usepackage{graphicx}
\usepackage{dcolumn}
\usepackage{bm}

\usepackage{color}
\usepackage{url}
\usepackage{verbatim}
\usepackage{lineno}
\usepackage[utf8]{inputenc}
\usepackage{color}
\usepackage{lineno}
\usepackage{mathtools}
\usepackage{orcidlink}
\usepackage{enumitem}

\DeclareMathOperator{\erf}{erf} 

\begin{document}



\title{A simple tool for weighted averaging of inconsistent data sets} 



\author{M.~Trassinelli \orcidlink{0000-0003-4414-1801}}
\email[]{martino.trassinelli@cnrs.fr}
\affiliation {Institut des NanoSciences de Paris, CNRS, Sorbonne Université, F-75005 Paris, France}
\author{M.~Maxton \orcidlink{0009-0005-9076-3101}}
\affiliation {Institut des NanoSciences de Paris, CNRS, Sorbonne Université, F-75005 Paris, France}

\date{\today}

\begin{abstract}
The weighted average of inconsistent data is a common and tedious problem that many scientists have encountered.
The standard weighted average is not recommended for these cases, and various alternative methods have been proposed. 
These approaches vary in suitability depending on the nature of the data, which can make selecting the appropriate method difficult without expertise in metrology or statistics.
For the analysis of simple data sets presenting inconsistencies, we discuss the method proposed by Sivia in 1996 based on Bayesian statistics.
This choice has the intention of maintaining generality while minimising the number of assumptions.
In this approach, the uncertainty associated with each input value is considered to be just a lower bound of the true unknown uncertainty.
The resulting likelihood function is no longer Gaussian but has smoothly decreasing wings, which allows for a better treatment of scattered data and outliers.  
To demonstrate the robustness and the generality of the method, we apply it to a series of critical data sets: simulations, CODATA recommended values of the Newtonian gravitational constant, and some particle properties from the Particle Data Group, including the proton charge radius.
A freely available Python library is also provided for a simple implementation of the proposed averaging method.
\end{abstract}


\maketitle

\section{Introduction}

The standard method for combining different independent evaluations $x_i$ of the same quantity is to use the weighted average 
\begin{equation}
	\label{eq:swa}
	\hat \mu  = \cfrac{\sum_i x_i/ \sigma^2_i}{\sum_i 1/ \sigma^2_i},
\end{equation}
that employs the inverse of the square of the associated uncertainties $\sigma_i$ as weights. The corresponding uncertainty is given by
\begin{equation}
	\label{eq:swa_unc}
	\sigma_{\hat  \mu}  = \sqrt{\cfrac{1}{\sum_i 1/ \sigma^2_i}}.
\end{equation}

The big advantage of such a procedure is the analy\-tical and simple formula that anyone can easily apply to any data set.
In addition, it is statistically well justified with a very small number of simple assumptions.
More importantly, the method is sufficiently universal to be considered as a standard procedure in the scientific community and can be found in any basic data analysis lecture.

However, the inverse-invariance method, referred to in the following pages simply as \emph{standard}, has a drawback.
As we can see from Eq.~\eqref{eq:swa_unc}, the final uncertainty depends only on the data uncertainties $\sigma_i$, but not on the data spread, which could be larger than the values of $\sigma_i$ (see, e.g., \cite{vonderLinden} for a more detailed discussion).
This is, however, a common scenario in science, possibly caused by an uncontrolled systematic effect in the measurement procedure or by different biases in measurements conducted in different laboratories and/or with different methods. 
Common questions that arise are how to take into account such information on the data dispersion in the calculation of a weighted average and how to treat outliers.

To answer such questions, several approaches have been proposed in the literature. 
A very common and basic method is to use the standard weighted average while artificially increasing its associated uncertainty.
But how should one choose objectively the uncertainty expansion factor?
The most common method was proposed by Birge \cite{Birge1932} almost one hundred years ago.
It is based on the $\chi^2$ value obtained by the difference between the standard weighted average and the single input values.
The uncertainty expansion factor $R_\mathrm{Birge}$, the \emph{Birge ratio}, is applied to the single uncertainties $ \tilde \sigma_i = R_\mathrm{Birge} \sigma_i$, with
\begin{equation}
R_\mathrm{Birge} = \sqrt{ \frac{1}{n-1} \sum_i \frac{(x_i - \hat \mu)^2}{\sigma_i^2}} = \sqrt{ \frac{\chi^2}{n-1} },
\end{equation}
where $n$ is the number of $x_i$ data points.
In this way, the final value of the reduced $\chi^2$ is adjusted to be close to unity, as expected for consistent data sets.
A modified version of the Birge ratio has been proposed in past works based on Bayesian statistics.
The scaling factor $R$ between the estimated uncertainty $\sigma_i$ and the real uncertainty $\sigma_i'$ is considered unknown but common to all data points \cite{Lira2007,Toman2012,Bodnar2014,vonderLinden}.
Assuming this, along with a reasonable prior on the value of $R$, the modified expansion factor
$R_\mathrm{Bayes} = \sqrt{(n-1)/(n-3)}R_\mathrm{Birge}$ is obtained.
Variations of this approach are discussed in Refs.~\cite{Mana2012,Bodnar2016}.

Because of the common scaling factor for each measurement result, the Birge ratio and its modified versions discussed above are however not well suited for interlaboratory averages, where very different systematic effects can occur.
To compensate partially for such an issue, past works \cite{Lira2007,Bodnar2016} proposed to assign a random bias $\beta_i$ to each measurement, with a common mean value and standard deviation $\sigma_\mathrm{bias}$ for the entire ensemble of measurements.
Here, a double marginalisation over the $\beta_i$ values and their shared uncertainty $\sigma_\mathrm{bias}$ is required.
An evolution of such an approach has been proposed \cite{Rukhin2019}.

Many other approaches exist (see, e. g., \cite{vonderLinden, Dose2003, Possolo2023} for a general overview).
Like the methods discussed above, they implicitly assume that the uncertainty $\sigma_i$ is a lower bound of the real uncertainty $\sigma_i'$.
This simple and clear statement has been translated into formulas by Sivia and Skilling \cite{Sivia1996b,Sivia}, avoiding common scaling factors (like $R_\mathrm{Birge}$) or random bias dispersion (like $\beta_\mathrm{bias}$), but considering for each point a modification of the Gaussian distribution by the marginalisation over $\sigma_i'$.
For this approach, a prior probability $p(\sigma_i')$ for $\sigma_i'$ has to be chosen.
The natural choice would be the non-informative Jeffreys' prior $p(\sigma_i') \propto 1 / \sigma_i'$.
This prior is indeed invariant over a change of coordinates for the parameter of interest with a minimization of possible biases.
Its implementation is especially interesting for scale parameters such as $\sigma_i$. (See e.g. Ref.~\cite{vonderLinden} for a detailed discussion.)
If not constrained by an upper bound, this choice causes divergence because of the non-integrability of the resulting final probability distribution associated with each datum.
To avoid this problem, a modified prior $p(\sigma_i') \propto 1 / (\sigma_i')^2$ has been proposed and discussed in Refs.~\cite{Sivia,Shirono2010,Mana2012}.
Alternatively, the inverse-gamma distribution, which is integrable but introduces additional parameters, is proposed as a prior distribution \cite{vonderLinden,Dose2003}.
Other more complex approaches with no lower bounds for $\sigma_i'$ can be found in Refs.~\cite{Sivia,vonderLinden}.

Note that alternative approaches based on non-Gaussian distributions have also been discussed in the literature.
In particular, the Student’s t-distribution is well known for its robustness to outliers \cite{Dawid1973, OHagan1979}.
Dose \cite{Dose2007} proposes using Laplace and hyperbolic cosine likelihoods in addition to the Gaussian likelihood and averaging the three distributions by applying Bayesian model selection.

Here, we discuss in detail the method by Sivia and Skilling \cite{Sivia1996b,Sivia}.
As will be seen, this approach is easily implementable while avoiding the issues associated with the standard weighted average.
Like the standard weighted average, basic assumptions are kept simple and very general.
To facilitate the application of the proposed method in daily data analysis, a Python code is provided.

Details on the derivation of the method are presented in Sec.~\ref{sec:derivation}.
In Section \ref{sec:tests}, the proposed method is compared to the other methods in a series of cases using simulated and real data.
In section \ref{sec:code}, the Python library based on the introduced method is presented. 
The final section is devoted to the discussion and conclusion.

\section{Derivation of the weighted average for inconsistent data}
\label{sec:derivation}

\subsection{General considerations}

The standard weighted average of independent measurements $x_i$ with uncertainties $\sigma_i$ is obtained by maximising the total probability $p(\mu | \{x_i,\sigma_i\})$ of the mean value $\mu$ for a given set of measured values $\{x_i,\sigma_i\}$.
Because of the independence of the measurements, it results from the product of the single probabilities for each measurement, with
\begin{equation}
\label{eq:wa_gen}
p(\mu | \{x_i,\sigma_i\}) \propto \prod_i p(x_i | \mu, \sigma_i) \ p(\mu).
\end{equation}
With the above formula, we implicitly assume that the probability distribution of each datum $x_i$ can be described by a distribution with only two parameters, its mean and standard deviation, and that the best estimation of the standard deviation is given by the provided values $\sigma_i$.
In the following sections, we will question the latter assumption.

When a Gaussian distribution is considered for each $x_i$ together with a flat prior probability for $\mu$ (a Jeffreys' prior), the most probable value $\hat \mu$ is given by the standard weighted average, i.e., Eq.~\eqref{eq:swa}.
The associated uncertainty $\sigma_{\hat \mu}$ given in Eq.~\eqref{eq:swa_unc} is simply derived by the uncertainty propagation in $\hat \mu = f(x_1, x_2, \ldots)$ of the uncertainty of the single $x_i$ values (see e.g. \cite{Bevington,Taylor}).

An alternative derivation of Eq.~\eqref{eq:swa_unc} can be obtained from the second derivative of the logarithm of $p(\mu | \{x_i,\sigma_i\})$ by supposing that the final probability distribution can be well approximated by a Gaussian distribution, where
\begin{equation}
	\label{eq:sigma_mu_gen}
	\sigma_{\hat \mu} = \left( - \frac{\partial^2 }{\partial \mu^2} \log [p(\mu | \{x_i,\sigma_i\}) ] \bigg|_{\mu = \hat \mu} \right)^{-\frac{1}{2}}.
\end{equation}

In line with standard methods, we consider the average as the value $\hat \mu$ that maximises Eq.~\eqref{eq:wa_gen} (assuming the Jeffreys' prior $p(\mu) = \text{const.}$), with its uncertainty given by Eq.~\eqref{eq:sigma_mu_gen}.
The single probabilities $p(x_i | \mu, \sigma_i)$ are no longer Gaussian, but are instead obtained by assuming certain hypotheses on the priors and performing marginalisations.

As stated above, our ultimate goal is to provide a tool for obtaining a robust weighted average applicable to a broad range of cases, which can be easily used as an alternative to the standard inverse-variance weighted average.
Two prerogatives are thus essential: to propose something very general and to consider a minimal number of assumptions.
For these purposes, we adopt the pessimistic framework where the uncertainty $\sigma_i$ is regarded as a lower bound of the real uncertainty $\sigma'_i$, without any additional assumptions on the possible biases and relations influencing the available data set. 
Any systematic error is considered to be included in the unknown uncertainty $\sigma_i'$.

Because of the unknown value of $\sigma'_i$ of each measurement $x_i$, the associated probability distribution is obtained by marginalising over $\sigma'_i$:
\begin{equation}
	\label{eq:prob_sigma}
	p(x_i | \mu, \sigma_i) = \int_{\sigma_i}^\infty p(x_i | \mu, \sigma'_i) p(\sigma'_i | \sigma_i)  \mathrm{d} \sigma'_i.
\end{equation}
If only pairs $(x_i, \sigma_i)$ of measured values and associated uncertainties are available, a Gaussian distribution $p(x_i | \mu, \sigma_i)$ can reasonably be assumed for each datum.
A choice for the prior probability distribution $p(\sigma'_i | \sigma_i)$ for $\sigma'_i$ has to be made.
The natural choice is a Jeffreys' prior, which is a non-informative prior that is invariant under reparametrisation to avoid introducing other possible biases, with
\begin{equation}
	\label{eq:Jeffreys}
	p(\sigma'_i | \sigma_i) = \begin{cases}
		\cfrac 1 {\log(\sigma^\mathrm{max}_i/\sigma_i)} \cfrac 1 {\sigma_i'} &\text{\quad for }\  \sigma_i \le \sigma'_i \le \sigma^\mathrm{max}_i,  \\
		0 &\text{\quad otherwise.} \\
	\end{cases}
\end{equation} 
The problem of such a prior is the introduction of an additional parameter $\sigma^\mathrm{max}_i$ for each data point.

\subsection{Sivia and Skilling's conservative weighted average}
\label{sec:conservative}

\begin{figure}
\begin{center} 
\includegraphics[width=0.5\textwidth]{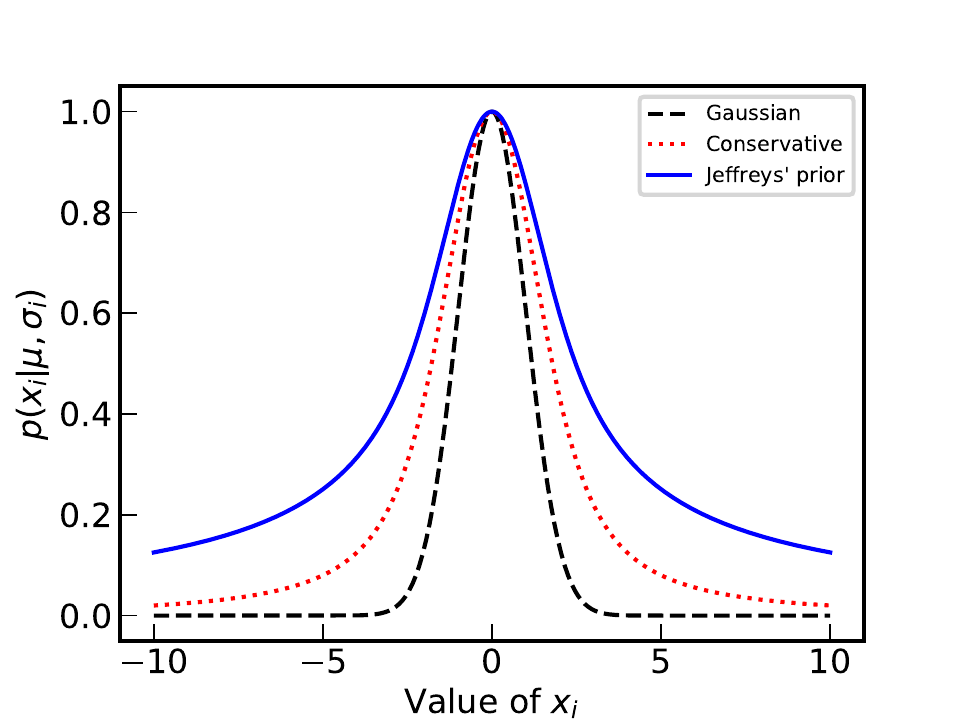}
\caption{Comparison between the different assumed probability distribution for each datum for $\mu=0,\sigma_i=1$.}
\label{fig:profiles}
\end{center}
\end{figure}

To avoid the introduction of additional parameters $\sigma^\mathrm{max}_i$ and keep a proper probability distribution for $p(\sigma'_i | \sigma_i)$, an alternative prior has been proposed by Sivia and Skilling \cite{Sivia}:
\begin{equation}
	\label{eq:prior_c}
	p(\sigma'_i | \sigma_i) = \frac {\sigma_i} {(\sigma'_i)^2}.
\end{equation}
We will refer in the following section to this approach as the \emph{conservative} approach, consistent with the authors’ own terminology.
Keeping the assumption of a Gaussian distribution for $p(x_i | \mu, \sigma'_i)$ and combining Eqs.~\eqref{eq:prob_sigma} and ~\eqref{eq:prior_c}, we obtain
\begin{equation}
	\label{eq:pxi_c}
	p(x_i | \mu, \sigma_i) = \frac {\sigma_i} {\sqrt{2 \pi}} \left[ \frac {1 - e^{-\frac{(x_i-\mu)^2}{2 \sigma^2_i}}}{(x_i-\mu)^2} \right].
\end{equation}
Compared with a Gaussian distribution, as visible in Fig.~\ref{fig:profiles}, the above expression is characterised by a significantly larger spread, with tails proportional to $1/(x_i)^2$.
Once plugged into Eq.~\eqref{eq:wa_gen}, such slowly descending tails supply sufficient flexibility to be tolerant of inconsistent data.
Unlike the standard weighted average, the maximising value $\hat \mu$ and its associated uncertainty $\sigma_{\hat \mu}$ have no analytical form, but can be easily determined with numerical methods. Now, both $\hat \mu$ and $\sigma_{\hat \mu}$ depend on the data spread.
Note that, because of the presence of the tails, even for consistent data sets, the final uncertainty of the weighted average is generally greater than the one obtained by the standard method.

\subsection{Limit solution with Jeffreys' prior}

A criticism that could be directed at Eq.~\eqref{eq:pxi_c} is that the choice of the prior probability distribution of $\sigma_i'$ in Eq.~\eqref{eq:prior_c} does not respect the non-informative criterion of Jeffreys' prior.
A possible solution to keep Jeffreys' prior without introducing any new parameters is given by Sivia in Ref.~\cite{Sivia1996b}. 
It considers the limit case $\sigma^\mathrm{max}_i \to \infty$.
While this approach can be considered even more conservative than the previous one, we will refer to it in the following section as \emph{Jeffreys' prior} approach.

When the Jeffreys' prior from Eq.~\eqref{eq:Jeffreys} is adopted and $\sigma^\mathrm{max}_i$ have a finite value, Eq.~\eqref{eq:prob_sigma} becomes
\begin{equation}
	p(x_i | \mu, \sigma_i) = \cfrac 1 {\log(\sigma^\mathrm{max}_i/\sigma_i)}  \frac{ \erf \left( \frac{x_i - \mu}{\sqrt{2} \sigma_i } \right) -\erf \left( \frac{x_i - \mu}{\sqrt{2} \sigma^\mathrm{max}_i } \right)}{2 (x_i - \mu)}.
\end{equation}
Compared to Eq.~\eqref{eq:pxi_c}, we can note that the distribution tails decrease even more smoothly than in the conservative approach, with a dependency of $1/x_i$ instead of $1/x_i^2$, which can tolerate the presence of outliers even better.
These more pronounced tails are well visible in Fig.~\ref{fig:profiles}.
For each individual datum, $p(x_i | \mu, \sigma_i)$ is an improper probability distribution because it is non-integrable. 
However, when at least two data values are considered, the probability $p(\mu | \{x_i,\sigma_i\})$ (Eq.~\eqref{eq:wa_gen}) is well defined, and the weighted average can be determined without ambiguities.

The logarithmic form of the total probability $p(\mu | \{x_i,\sigma_i, \sigma^\mathrm{max}_i \})$ is given by
\begin{multline}
	\log [p(\mu | \{x_i,\sigma_i, \sigma^\mathrm{max}_i \})]  \\
	= \sum_i  \log \left[ \frac{ \erf \left( \frac{x_i - \mu}{\sqrt{2} \sigma_i } \right) -\erf \left( \frac{x_i - \mu}{\sqrt{2} \sigma^\mathrm{max}_i } \right)}{2 (x_i - \mu)} \right] - C,
\end{multline}
where there is now a dependency on $\{ \sigma^\mathrm{max}_i \}$, and
\begin{equation}
	C = \sum_i \log \left[ \log (\sigma^\mathrm{max}_i/\sigma_i) \right].
\end{equation}
This constant term does not play a role in the search for $\hat \mu$, as $C$ depends only on the boundary values  $\sigma_i$ and $\sigma_i^\mathrm{max}$.

The limit $\sigma^\mathrm{max}_i \to \infty$ of the above equation is
\begin{multline}
	\log [p(\mu | \{x_i,\sigma_i\})] = \lim_{ \sigma^\mathrm{max}_i  \to  \infty  } \log [p(\mu | \{x_i,\sigma_i, \sigma^\mathrm{max}_i \})]  \\
	= \sum_i  \log \left[ \frac{ \erf \left( \frac{x_i - \mu}{\sqrt{2} \sigma_i } \right) }{2 (x_i - \mu)} \right]  - C^\infty,
\end{multline}
where the constant $C^\infty = \lim_{ \sigma^\mathrm{max}_i  \to  \infty  } C = \infty$ is indeed divergent, but for more than one data point, the distribution is integrable, and the variance is finite for more than two data points.
In particular, the position of the maximum and the value of the second derivative and thus the weighted average and its uncertainty are well defined.
As in the case of the conservative weighted average, no analytical solution is available for $\hat \mu$ and $\sigma_{\hat  \mu}$, so the solution must be found numerically.

In the following section, we will refer to this approach as the \emph{Jeffreys' weighted average}.

\section{Some applications}
\label{sec:tests}

In this section, we present a series of applications for common data analysis cases to test the reliability of the conservative and Jeffreys' weighted average. 
In the first subsection, we will study simulated data with known theoretical values of mean and standard deviation.
In the second subsection, an analysis of the different values of the Newtonian gravitational constant from past CODATA compilations is proposed.
The third subsection is dedicated to fundamental particle properties, including the controversial data set of the proton charge radius.
Additional implementations of the Jeffreys' weighted average for inconsistent data measured using the same experimental apparatus can be found in Ref.~\cite{Duval2024}.

\subsection{Synthetic tests}\label{sec:synthetic}

\begin{figure}
\begin{center} 
\includegraphics[clip, trim = 0 40 40 70, width=0.5\textwidth]{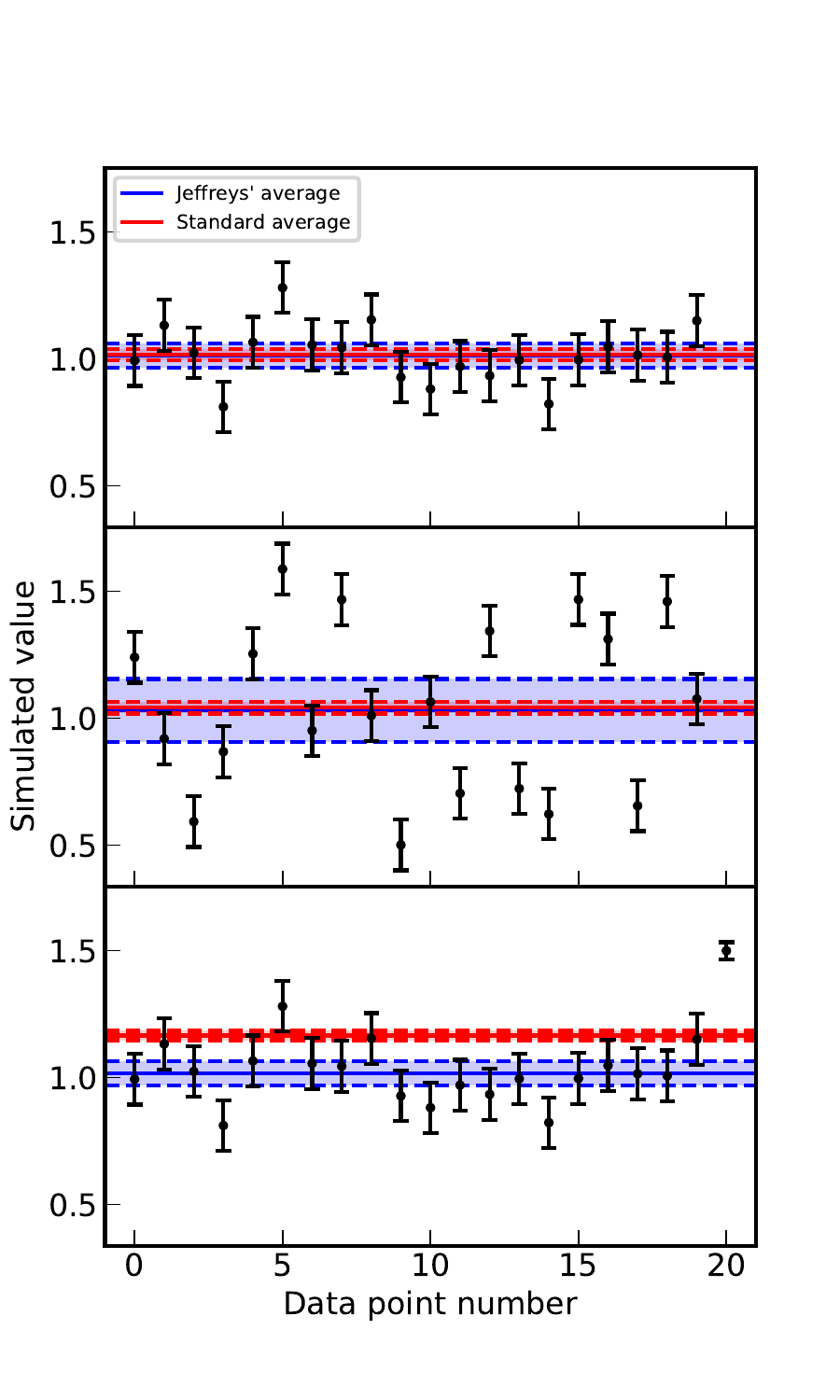}
\caption{Standard and Jeffreys' weighted averages of different simulated data sets: data randomly sampled from a normal distribution (top) and with the addition of a random bias (middle) or an outlier (bottom).}
\label{fig:synthetic}
\end{center}
\end{figure}

Different simulated data sets are considered for comparing the different averaging methods:
\begin{enumerate}

\item The first set of $N=20$ values $x_i$ is obtained by a normal distribution with a mean value of $\mu=1$ and a standard deviation equal to $\sigma=0.1$.
For each data point, the uncertainty $\sigma_i = \sigma$ is considered.

\item The second set simulates inconsistent data. It is derived from set 1 by adding a random bias, with a standard deviation of $\sigma_\mathrm{bias} = 10\,\sigma$ and a mean of $\mu_\mathrm{bias} = 0$, to each data point.

\item The third set is the same as the first but with the addition of an outlier at $\mu +5 \,\sigma$ that has an uncertainty of $\sigma_\mathrm{out} = \sigma/3$ with a total number of $N=21$ values.
\end{enumerate}
The choice of $\sigma_\mathrm{bias} = 10\,\sigma$ for set 2 reflects some typical measurement scenarios, such as those reported in Ref.~\cite{Duval2024}.
For set 3, the $5 \sigma$ separation is motivated by the five-sigma threshold usually implemented in particle physics.
The value of $\sigma_\mathrm{out} = \sigma/3$ has been chosen arbitrarily to illustrate a clear deviation, but is still less extreme than the measurement discussed in Sec.~\ref{sec:G}. 
The values of the sets are provided in the \hyperref[supp-data]{Supplementary data}.

\begin{table}
\caption{\label{tab:synthetic_1} Weighted average values with corresponding uncertainties for the synthetic data set 1 (data shown in Fig.~\ref{fig:synthetic}, top).}
\begin{ruledtabular}
\begin{tabular}{l l l}
Weighted average type & Average $\hat \mu$ & Unc. $\sigma_{\hat \mu}$\\
\hline
Inverse-variance (standard) &  1.015  &  0.022  \\
Inverse-variance with Birge ratio &  1.015  &  0.025$^a$  \\
Conservative &  1.013  &  0.037  \\
Jeffreys' prior  &  1.013  &  0.047  \\
\end{tabular}
\end{ruledtabular}
      \footnotesize \raggedright
      $^a$Birge ratio equal to 1.1
\end{table}

\begin{table}
\caption{\label{tab:synthetic_2} Weighted average values with corresponding uncertainties for the synthetic data set 2 (data shown in Fig.~\ref{fig:synthetic}, middle).}
\begin{ruledtabular}
\begin{tabular}{l l l}
Weighted average type & Average $\hat \mu$ & Unc. $\sigma_{\hat \mu}$\\
\hline
Inverse-variance (standard) &  1.041  &  0.022  \\
Inverse-variance with Birge ratio &  1.041  &  0.076$^a$  \\
Conservative &  1.04  &  0.094  \\
Jeffreys' prior  &  1.031  &  0.124  \\
\end{tabular}
\end{ruledtabular}
      \footnotesize \raggedright
      $^a$Birge ratio equal to 3.4
\end{table}

\begin{table}
\caption{\label{tab:synthetic_3} Weighted average values with corresponding uncertainties for the synthetic data set 3 (data shown in Fig.~\ref{fig:synthetic}, bottom, detailed analysis in Fig.~\ref{fig:synthetic_out}).}
\begin{ruledtabular}
\begin{tabular}{l l l}
Weighted average type & Average $\hat \mu$ & Unc. $\sigma_{\hat \mu}$\\
\hline
Inverse-variance (standard) &  1.166  &  0.019  \\
Inverse-variance with Birge ratio  &  1.166  &  0.054$^a$  \\
Conservative &  1.019  &  0.037  \\
Jeffreys' prior  &  1.018  &  0.047  \\
\end{tabular}
\end{ruledtabular}
      \footnotesize \raggedright
      $^a$Birge ratio equal to 2.9
\end{table}

\begin{figure}
\begin{center} 
\includegraphics[width=0.5\textwidth]{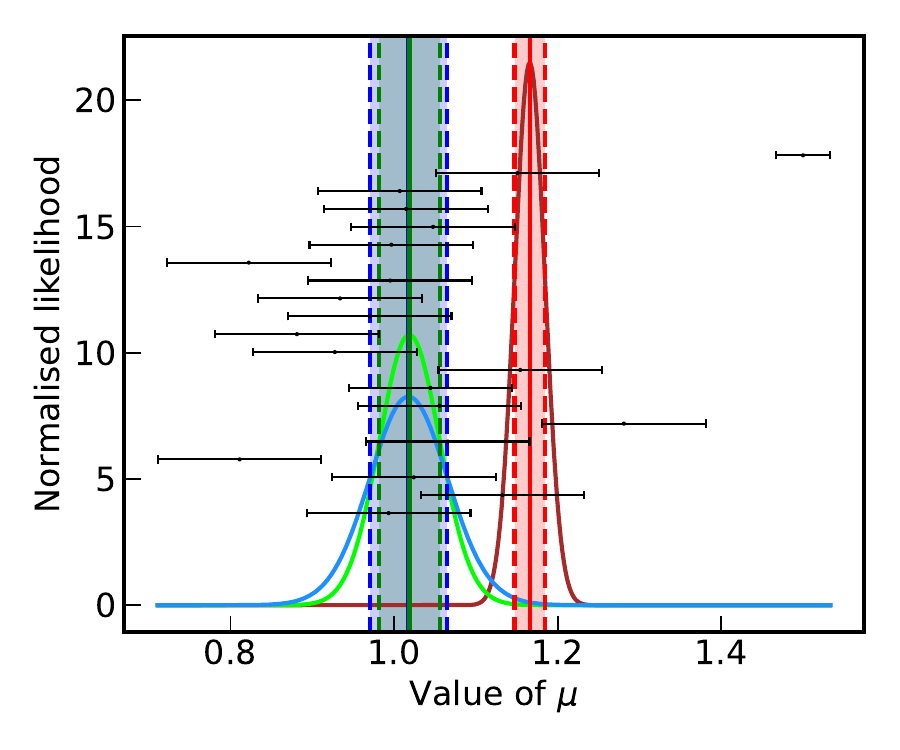}
\caption{Final likelihood distributions for the data set 3 together with the input values (black).
The solid lines represent the standard (red), conservative (green) and Jeffreys' prior likelihood distribution (blue).
The corresponding average and standard deviation are marked by vertical lines.}
\label{fig:synthetic_out}
\end{center}
\end{figure}

For the three sets, the standard weighted average, with or without Birge ratio correction, is compared to the conservative and Jeffreys' weighted averages.
The results are presented in Tables~\ref{tab:synthetic_1}--\ref{tab:synthetic_3} and in Fig.~\ref{fig:synthetic} (for the standard and Jeffreys' weighted averages only) and Fig.~\ref{fig:synthetic_out}.
As we can see, for normal and inconsistent data without outliers, the mean value is well reproduced by all methods.
Because of the pessimistic priors on $\sigma_i'$, the Jeffreys' and conservative final uncertainties are generally larger than the standard uncertainty. For consistent data (set 1), they are larger by a factor of about two.
As expected, for the inconsistent data (set 2), the uncertainty associated with the standard weighted average is significantly smaller than the others, with the Jeffreys' uncertainty being the largest, followed by the conservative and the Birge uncertainties.
Jeffreys’ weighted average uncertainty is almost six times larger than the standard weighted average uncertainty and double that of the Birge-ratio-corrected uncertainty.

When an outlier is present, Jeffreys' weighted average values are quite different from the standard weighted average. 
The effect of the presence of an outlier (set 3) is clearly visible in Fig.~\ref{fig:synthetic_out}, where the final likelihoods are plotted together with the data, and in the results presented in Table \ref{tab:synthetic_3}.
As we can see, the effect on the standard likelihood, obtained by the product of Gaussian distributions, is drastic, with a shift in the direction of the outlier. 
If the data uncertainties are regarded as lower bounds only, the effect is greatly mitigated, resulting in just an asymmetry of the tails for the Jeffreys' and conservative priors, which have very similar final probability distributions.
This behaviour looks quite similar to other methods that utilize the deviation of the data point from the calculated value $\hat \mu$ to estimate the possibility of missing uncertainty contributions \cite{Lira2007,Huang2018}.
However, unlike these past works, here the distributions are derived from the initial assumptions on the uncertainties, for which we consider \emph{a priori} that the provided values $\sigma_i$ are only lower bounds of the real uncertainty.

\subsection{The Newtonian constant of gravitation}
\label{sec:G}

A significant example of an average between independent and possibly inconsistent measurements is the determination of the Newtonian constant of gravitation, which, due to the difficulties associated with its measurement, has long been the fundamental constant with the highest relative uncertainty.
Such difficulties are mainly due to the challenging experimental conditions, where very small forces must be isolated from a noisy environment \cite{Rothleitner2017,Merkatas2019}.
The official value is provided by CODATA with a standard inverse-variance weighted average, and the associated uncertainty is eventually multiplied by an expansion factor to maintain consistency between the final result and the considered measurements.

\begin{figure*}
\begin{center} 
\includegraphics[width=0.85\textwidth]{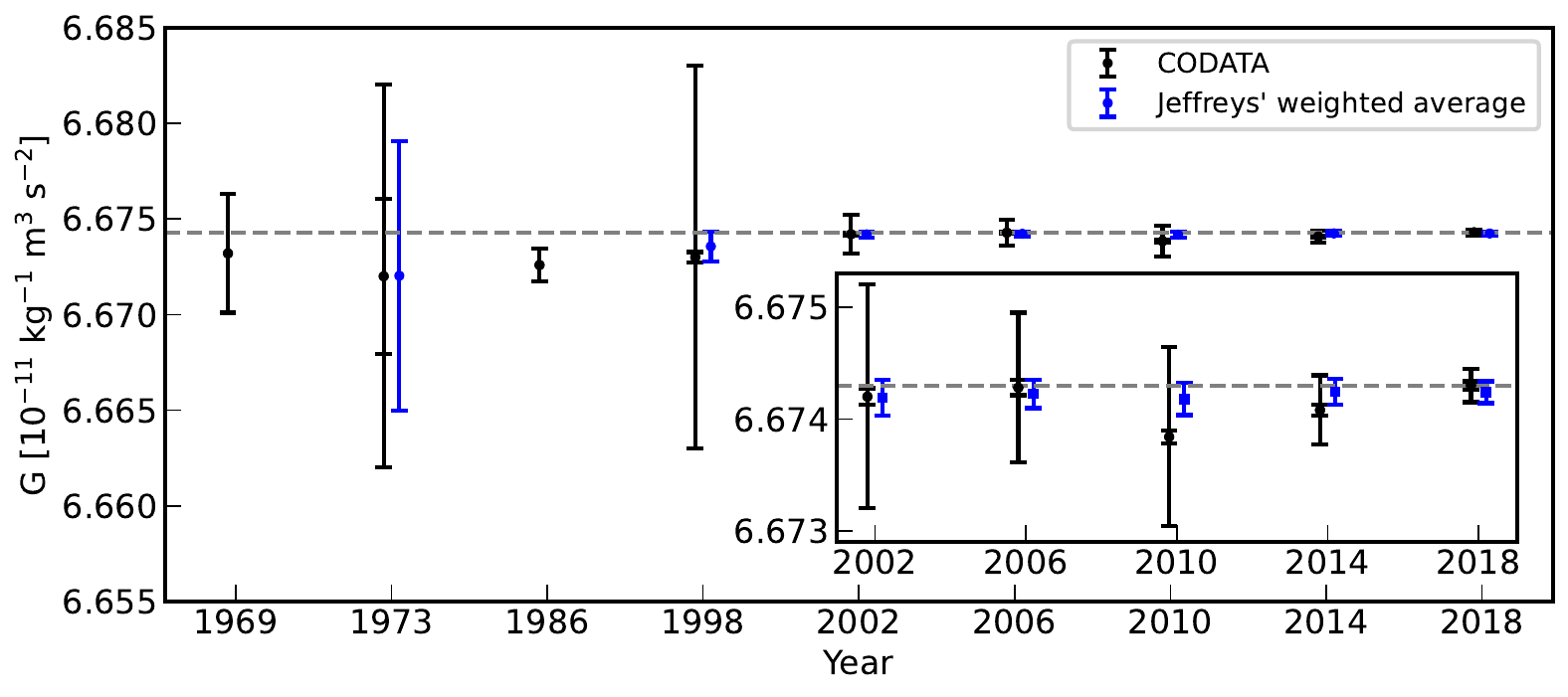}
\caption{Comparison between the official CODATA values \cite{CODATA1969,CODATA1973,CODATA1986,CODATA1998,CODATA2002,CODATA2006,CODATA2010,CODATA2014,CODATA2018} of the Newtonian constant and the values obtained by the Bayesian weighted average using Jeffreys' prior.
CODATA values obtained from single measurements are presented alone, as no weighted average could be performed.
The small error bar of the CODATA values indicates the uncertainty calculated by the standard weighted average, and the large one indicates the recommended uncertainty.
The horizontal dashed line corresponds to the latest CODATA value (2022 edition \cite{CODATA2022}, equal to the 2018 edition value).}
\label{fig:CODATA_G}
\end{center}
\end{figure*}

Here, we apply the Jeffreys' weighted average to all data sets included in the different editions of the CODATA compilation \cite{CODATA1969,CODATA1973,CODATA1986,CODATA1998,CODATA2002,CODATA2006,CODATA2010,CODATA2014,CODATA2018,CODATA2022}.
The results are presented together with the official values in Fig.~\ref{fig:CODATA_G}.
For each reported CODATA value, the large error bar corresponds to the recommended value of the uncertainty, and the small one to the uncertainty calculated by the standard weighted average.
As we can see, the standard weighted average is, for some years, several standard deviations away from the most recent CODATA value from 2018 (the horizontal dashed line in the figure, which is the same as the more recent CODATA 2022 value \cite{CODATA2022}), considered here as the reference value.
Contrary to standard procedures, one can see that the values obtained by the Jeffreys' weighted average are consistently in good agreement, being less than one standard deviation away from the most recent CODATA value, and are characterised by a more plausible uncertainty.

\begin{table}
\caption{\label{tab:codata}Weighted average values with corresponding uncertainties for the data set of the CODATA 1998 compilation for the Newtonian gravitational constant and corresponding recommended value.
All reported values are expressed in units of $10^{-11}$~m$^3$ kg$^{-1}$ s$^{-2}$.}
\begin{ruledtabular}
\begin{tabular}{l l l}
Weighted average type & Average $\hat \mu$ & Unc. $\sigma_{\hat \mu}$\\
\hline
Inverse-variance (standard) &  6.6827  &  0.0003  \\
Inverse-variance with Birge ratio &  6.6827  &  0.0063$^a$   \\
Conservative &  6.6735  &  0.0006  \\
Dose model average \cite{Dose2007} &  6.6746 & 0.0035 \\
Jeffreys' prior  &  6.6736  &  0.0008  \\
\hline
CODATA 1998 \cite{CODATA1998}  & 6.673 &0.010$^b$  \\
\end{tabular}
\end{ruledtabular}
      \footnotesize \raggedright
      $^a$Birge ratio equal to 23.6\\
      $^b$Corresponding scale factor equal to 37
\end{table}

\begin{figure}
\begin{center} 
\includegraphics[width=0.5\textwidth]{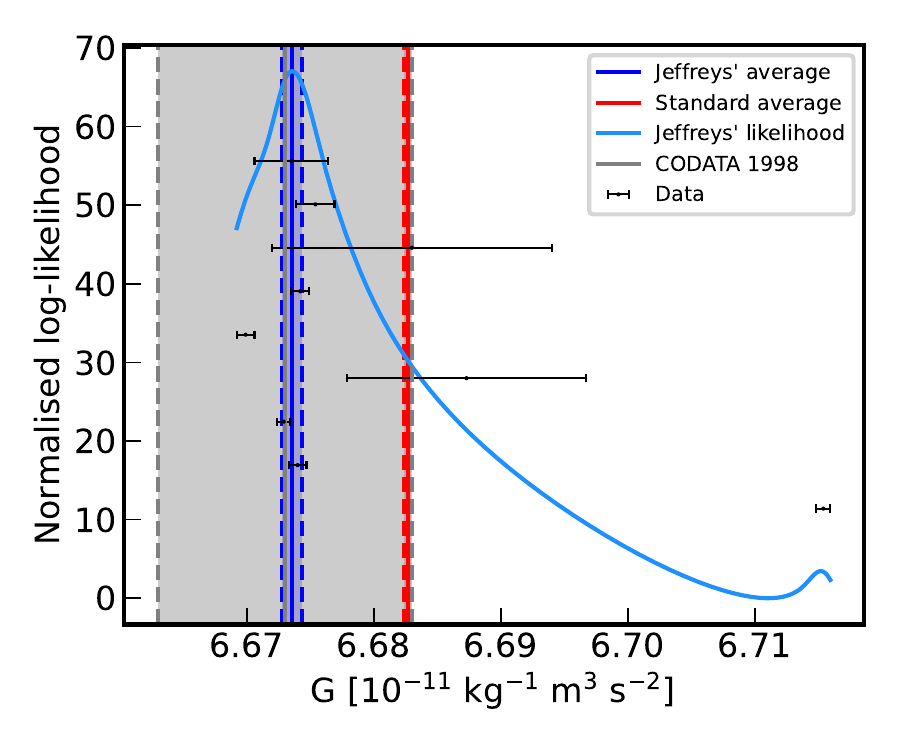}
\caption{Final likelihood distribution (in log scale) of the measurements of the Newtonian constant included in the CODATA 1998 compilation \cite{CODATA1998}. The CODATA 1998 recommended value is also reported (in grey), which differs from the standard weighted average for the considered measurements (in red).}
\label{fig:CODATA_G_1998}
\end{center}
\end{figure}

The 1998 case is particularly difficult due to the inconsistency within the data set, arising from one very precise measurement \cite{Michaelis1995} that differs significantly from the average of the other measurements.
The value was later found to be affected by a large systematic error \cite{Michaelis2004}. 
Details of the analysis of this specific case are presented in Table~\ref{tab:codata} and Fig.~\ref{fig:CODATA_G_1998}. 
The CODATA recommended value is obtained by the standard weighted average of all values, excluding the suspicious measurement. 
The corresponding uncertainty is obtained by applying an expansion factor of 37 to the standard weighted average uncertainty to reflect the presence of the outlier. 
More precisely, the final uncertainty has been chosen to ensure that the difference between the recommended value and the outlier is four times larger than the final uncertainty.
Like in set 3 of the previous section, the Jeffreys' weighted average is only slightly affected by the outlier in this challenging case.
Looking at Table~\ref{tab:codata}, one can note that the Jeffreys' weighted average has a significantly smaller uncertainty than the method tested on the same data set by Dose \cite{Dose2007}.
There, a Bayesian model average was applied on different likelihood functions.

\subsection{Particle properties}
\label{sec:PDG}

Another field that has to deal with very different measurements to compile reference data is particle physics.
The Particle Data Group (PDG) \cite{PDG2024}, which is providing the official reference values, implements a very well-documented procedure, mainly based on the Birge ratio and data selection.
The goal of this selection is to minimise possible correlations between considered data and to exclude evident outliers.
The muon magnetic anomalous moment is an exception.
No measurements are selected except the one considered as unique reference from the most recent experiment \cite{Aguillard2023} (which is now in agreement with the Standard Model predictions \cite{Ignatov2024}). 

\begin{figure*}
\begin{center} 
\includegraphics[width=0.85\textwidth]{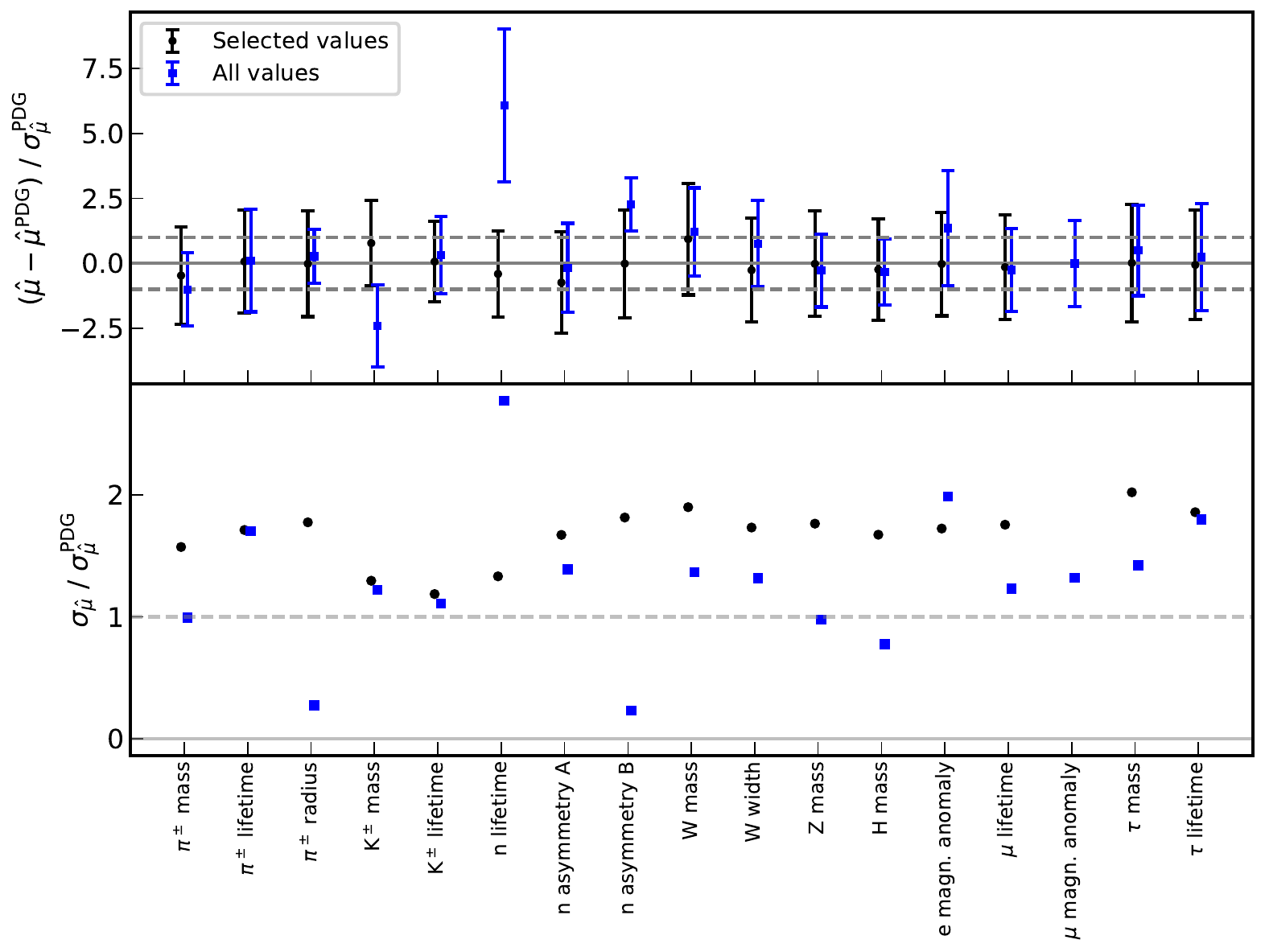}
\caption{Top: Comparison between the values obtained by the Bayesian weighted average using Jeffreys' prior and the official PDG values of several particle properties  \cite{PDG2024}.
All values are normalised to the PDG uncertainties.
The two dashed horizontal lines correspond to $\hat \mu - \hat \mu^\mathrm{PDG} = \pm \sigma_{\hat \mu}^\mathrm{PDG}$, i.e. to $\pm 1$ in the relative scale.
Bottom: Ratio between the uncertainties obtained for the Jeffreys' weighted average and the official PDG uncertainties.}
\label{fig:PDG}
\end{center}
\end{figure*}

In this section, we evaluate the Jeffreys' weighted average of some particle properties and compare the results with the PDG recommended values.
Moreover, to test the robustness of the Jeffreys' weighted average, we evaluate two sets: 
\begin{enumerate}[label=\Alph*:]

\item the measurements selected by the PDG for the determination of the recommended value.

\item the whole set of values listed by PDG.
\end{enumerate}
For both sets, no correlations between the data have been considered. 
This assumption is, however, not well adapted to set B, where strongly correlated measurements are present in some cases.

The obtained average values are compared with the PDG values in Fig.~\ref{fig:PDG}.
Because of the very different quantities, we normalise the difference of the Jeffreys' weighted average to the PDG value by the uncertainty provided by the PDG.
It comes as no surprise that, when the selected data are considered, the Jeffreys' weighted average is in good agreement with the PDG values.
Similar to the case of the simulated data sets, the associated final uncertainty is generally larger than the PDG uncertainty by at most a factor of 2.2.
For set B, smaller values of the final uncertainty can also be found because of the larger set of considered data.
For both sets, deviations of less than two standard deviations are observed, indicating the good robustness of the Bayesian method even in difficult cases. 
The exception of the deviation of K$^\pm$ meson mass for set B is caused by very strong correlations between measurements.
Moreover, the deviation of the neutron asymmetry parameter B is caused by a single additional value in set B, which is excluded by the PDG (set A).
Like the case of the Newtonian constant in CODATA 1998, this very precise additional value has a strong influence due to the lack of precise measurements. 

\begin{figure}
\begin{center} 
\includegraphics[width=0.5\textwidth]{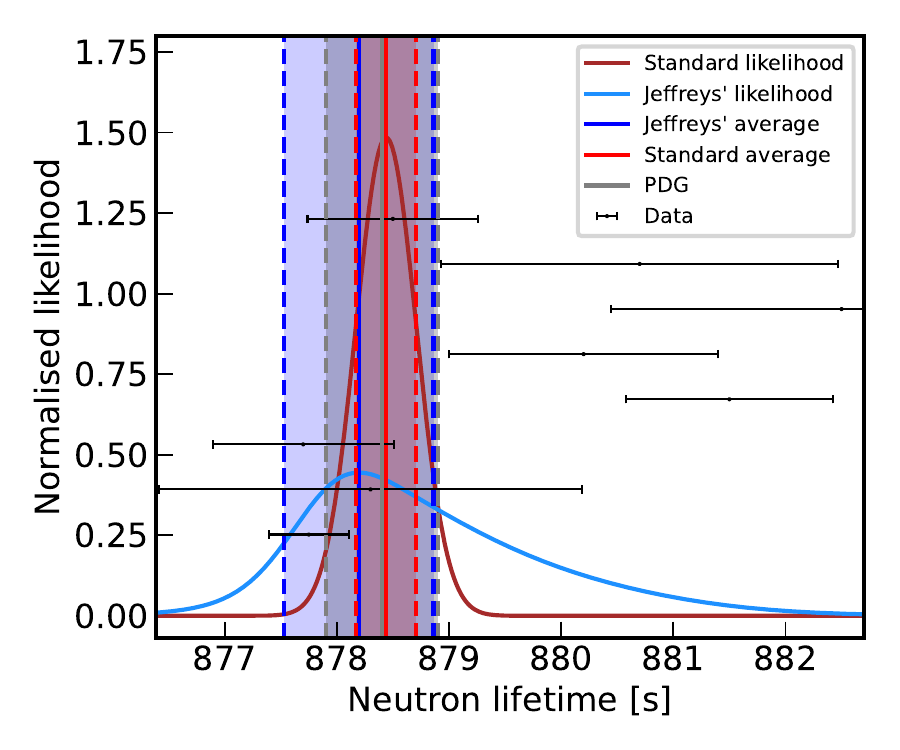}
\includegraphics[width=0.5\textwidth]{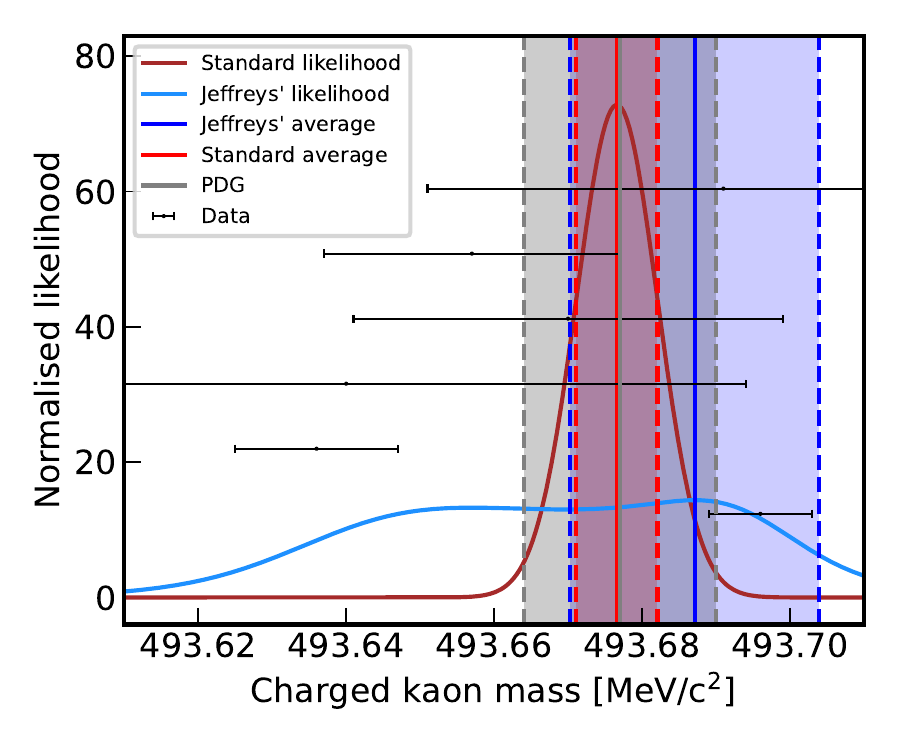}
\caption{Final likelihood distribution (in linear scale) corresponding to the PDG measurement selection for the neutron lifetime (top) and the charged kaon mass (bottom).
In these cases, the final probability distribution generated by the Bayesian method is highly asymmetric and cannot be approximated by a Gaussian distribution. Therefore, the whole probability distribution should be considered for further inference analyses instead of any kind of weighted averages.}
\label{fig:PDG_asymmetries}
\end{center}
\end{figure}

However, even if in good agreement with PDG values, some results must be treated with caution. 
In contrast to the standard weighted average, typically associated with a sharp (Gaussian) probability distribution, the probability distribution for the Jeffreys' prior may more readily exhibit strong asymmetry and multimodality.
As an example, the deviation in the neutron lifetime in set B is caused by an asymmetry in the final probability distribution.
Similar cases are found for the K$^\pm$ meson mass (set A only), the neutron lifetime (set A only), neutron asymmetry parameter A, muon mass, and the e$^-$ magnetic moment anomaly.
For the latter in particular, no deviation from the PDG value is visible.
Two typical examples are shown in Fig.~\ref{fig:PDG_asymmetries}, presenting the neutron lifetime and the charged kaon mass in detail.
As we can see, the use of any kind of weighted average is not appropriate because it does not reflect the final probability distribution, which should be considered for further inferences instead. 
These considerations are in complete agreement with the PDG recommendations that for these non-trivial cases point out possible issues with these sets and provide an ideogram corresponding to the combination of the measurement results (assumed to be Gaussian with a weight proportional to $1/\sigma_i$, and not to $1/\sigma_i^2$ like for the standard weighted average) to underline the importance of the single values in the average. 
Unlike the standard and PDG methods, such a conclusion can be directly deduced by looking at the final probability distribution for the Jeffreys' prior.

\begin{figure}
\begin{center} 
\includegraphics[width=0.5\textwidth]{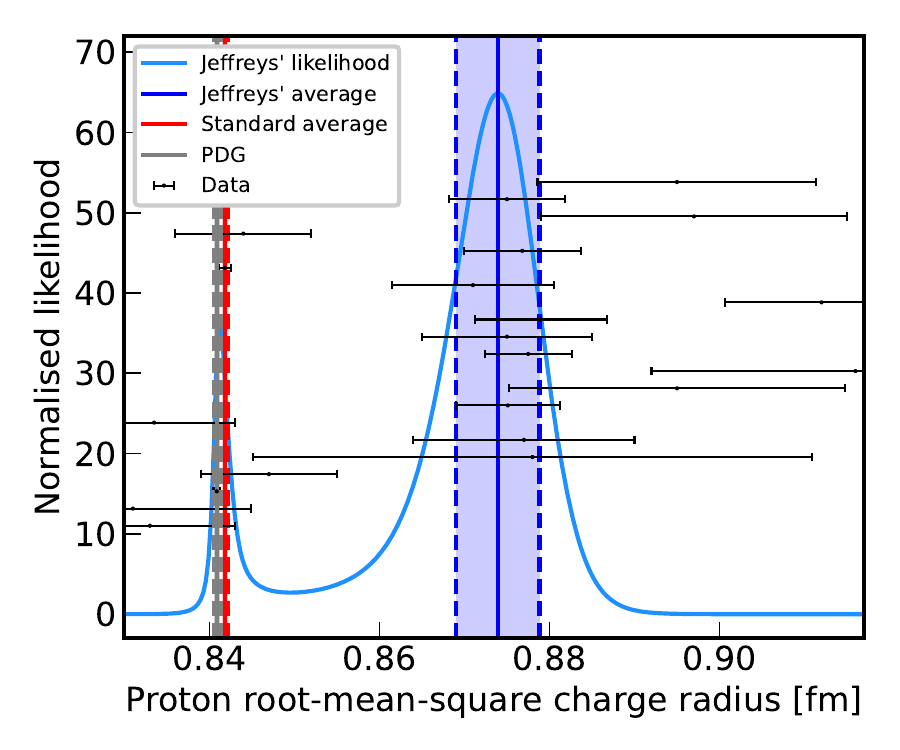}
\caption{Final likelihood distribution (in linear scale) corresponding to the PDG measurement selection for the proton radius. The PDG official value is also reported, which is very close to the standard weighted average.}
\label{fig:PDG_proton}
\end{center}
\end{figure}

An extreme example of asymmetry and multimodality in the final probability distribution is the case of the proton rms charge radius.
For this quantity, very different results are obtained from the Jeffreys' weighted average with respect to the recommended PDG value and the standard weighted average (so different that they are not reported in Fig.~\ref{fig:PDG}). 
As we can see from the plot of the probability distribution (Fig.~\ref{fig:PDG_proton}), a pronounced bi-modal distribution appears when evaluating the entire ensemble of available data.
In this situation, any weighted average, for which we always assume unimodality of the associated probability distribution, is not suitable.
This situation is similar to the case of a data set composed of only two data points with different values but the same uncertainties. Regardless of the chosen method, a weighted average will propose the midpoint between the data points due to the symmetry of the problem.
In such cases, as written above, the whole probability distribution should be taken into account as well as the critical judgement of experienced researchers.

\section{The associated code}
\label{sec:code}

Despite the well-known problems of the standard weighted average based on the inverse of the variance, its widespread use is undoubtedly largely due to the simplicity of its formula, which can be easily employed by anyone.
The Jeffreys' weighted average, as well as other alternative methods proposed in the literature, has the disadvantage of requiring numerical methods for determining the weighted average and its uncertainty due to the complexity of the associated analytical formulas.
The need for improved averaging methods is, however, prevalent, but due to the implementation difficulties, such alternative methods are generally quickly abandoned, sometimes in favour of the simpler Birge ratio.
To close this gap, we provide a numerical tool for the proposed weighted averaging method.
More precisely, we propose the Python library \textsc{bayesian\_average}, which can be easily installed in any Python environment using the \texttt{pip} command and is also freely available via the GitHub repository \footnote{\url{https://github.com/martinit18/bayesian_average}}.
In addition to providing the weighted average based on the Jeffreys' and conservative method priors, the standard inverse-variance method and its modified version with the Birge ratio are included for comparison. 
A graphical tool is available to plot the final weighted averages and the associated final likelihood probability distributions together with the input data.
Figures \ref{fig:synthetic_out}, \ref{fig:CODATA_G_1998}, \ref{fig:PDG_asymmetries} and \ref{fig:PDG_proton} are typical outputs from \textsc{bayesian\_average} (with minimal changes to the label of the axes). 

\section{Discussion and conclusions}

We present a robust method for averaging inconsistent data, originally proposed by Sivia \cite{Sivia1996b}.
We suggest using this method for the analysis of simple data sets exhibiting inconsistencies as an alternative to the standard weighted average based on the inverse of the variance.
In addition, we propose a Python package to easily calculate such a weighted average.

Compared with other similar methods previously discussed in the literature, the number of working hypotheses is kept to the minimal requirements. 
The weighted average formulated by Sivia \cite{Sivia1996b} and based on Bayesian statistics is implemented to avoid formulating complex hypotheses on the nature and behaviour of the unknown component of the true uncertainties.
For each data point, a Gaussian distribution is considered, but the provided uncertainty $\sigma_i$ is regarded as a lower bound of the true uncertainty value. 
Using Bayes' theorem and assuming a non-informative Jeffreys' prior for $\sigma_i'$, a new probability distribution is obtained by marginalising over $\sigma_i' \in [\sigma_i,\sigma_i^\mathrm{max}]$. 
The new arbitrary parameters $\sigma_i^\mathrm{max}$ are eliminated by looking at the asymptotic solution of the resulting weighted average $\hat \mu$. 

Because of the independent treatment for each datum uncertainty $\sigma_i$, the approaches presented in the previous sections are well adapted to interlaboratory studies.
Each uncertainty $\sigma_i$ may be underestimated due to unknown systematic errors specific to the corresponding laboratory.

A similar approach has been implemented by Mana \cite{Mana2021}, using a prior on $\sigma'_i$ as in Eq.~\eqref{eq:prior_c}. 
Unlike our proposed approach, Mana assumes that for some cases $\sigma_i = \sigma'_i$ holds.
To determine which data satisfy this assumption, the probability of the hypothesis (i. e., that $\sigma_i = \sigma'_i$ is valid) is evaluated by Bayesian model selection, based on the marginal likelihood (Bayesian evidence).
Related approaches are discussed in the literature \cite{Toman2012,Rukhin2019}, where an unknown bias is considered for clusters of data, thereby leading to an additional uncertainty being added to the original one.
Other similar methods are proposed in Refs. \cite{Shirono2010,Huang2018}.

All of the above-mentioned implementations imply, however, introduce significant complexity and may reduce the generality of their application.
In contrast, although more pessimistic (implicitly treating all laboratories with caution), our proposed method is much simpler and more generally applicable, while remaining well suited for data obtained from the same measurement apparatus.

The method based on the Jeffreys' prior is applied to a series of cases that show its reliability and robustness. 
For this purpose, simulated data, CODATA values of the Newtonian constant, as well as a series of particle property values from PDG are considered.

In particular, in the presence of outliers, the method has proven to be a very robust tool. 
In the case of the synthetic data set 3 and the average of the Newtonian gravitational constant reported in CODATA 1998, the effect of the outlier is minimized and the reference value is recovered without problems.
Nonetheless, the treatment of data containing outliers is always delicate.
As an example, in the case of the proton radius, the 2010 measurement using muonic hydrogen \cite{Pohl2010} was considered an outlier and thus excluded from the CODATA compilation for several years \cite{CODATA2010,CODATA2014}.
Over time, however, this ``anomalous'' experiment motivated new measurements that highlighted systematic effects in past measurements not based on muonic hydrogen, leading to its inclusion in the CODATA compilation \cite{CODATA2022}.
For such cases involving outliers, as best practice, we suggest clearly detailing the data analysis choices and justifications of the chosen analysis method.
This transparency ensures that these decisions are understandable
and allows for the possible application of alternative approaches as new information or methods arise.
For such cases, the standard weighted average method should, however, be always avoided, as it does not account for potential misestimation of uncertainties.

In the case of particle properties, different scenarios are encountered. 
For the largest part of the cases, the Jeffreys' weighted average reproduces very well the PDG recommended values, but with a slightly larger uncertainty.
In some cases, however, we show that a weighted average procedure should be taken with caution, and the entire probability distribution should be considered instead, which is in agreement with the PDG recommendations.
This is particularly true for the analysis of the latest compilation of the proton radius, which shows a pronounced multimodality in the corresponding final probability.
With the presented method, these difficult cases are easily identified by looking at the final probability distribution.

Once more, it is important to emphasise that the Jeffreys' weighted average should not be regarded as a substitute for more thorough, expert analysis.
The critical judgement of experienced researchers is irreplaceable.
However, this method serves as a valuable tool for obtaining simple weighted averages from inconsistent data, especially in cases where the commonly used standard weighted average may lead to misleading results.

The focus of future developments will be on incorporating correlations between the input data for the calculation of the weighted average.

\acknowledgements{
We would like to thank François Nez for the very useful discussions on CODATA values, Louis Duval for providing the motivation to start this work with his analysis of ``inconsistent'' data, Mark Plimmer for the careful reading of the manuscript, Andrew Charman for introducing us to Sivia's earlier work that confirmed our independent calculations, and Udo von Toussaint for his valuable suggestions.} 

\subsection*{Author contributions}
The original idea was formulated by M.T.
The analysis was mainly performed by M.M.
The manuscript was written by both authors.

\subsection*{Code availability}
The weighted averages were calculated using the publicly available code {\sc Bayesian\_average} from the author, which is accessible in the repository \url{https://github.com/martinit18/bayesian\_average}.

\subsection*{Data availability}
The datasets generated and/or analysed during the current study are available in the \hyperref[supp-data]{Supplementary data}.

\subsection*{Supplementary data}
\label{supp-data}

\begin{table}[h!]
\caption{\label{tab:synthetic_sets} Synthetic data sets used in Section \ref{sec:synthetic}. Each column corresponds to a different set.}
\begin{ruledtabular}
\begin{tabular}{ccc}
\textbf{Set 1} & \textbf{Set 2} & \textbf{Set 3} \\ \hline
0.9934 $\pm$ 0.1000 & 1.2401 $\pm$ 0.1000 & 0.9934 $\pm$ 0.1000 \\
1.1324 $\pm$ 0.1000 & 0.9191 $\pm$ 0.1000 & 1.1324 $\pm$ 0.1000 \\
1.0241 $\pm$ 0.1000 & 0.5924 $\pm$ 0.1000 & 1.0241 $\pm$ 0.1000 \\
0.8112 $\pm$ 0.1000 & 0.8680 $\pm$ 0.1000 & 0.8112 $\pm$ 0.1000 \\
1.0656 $\pm$ 0.1000 & 1.2541 $\pm$ 0.1000 & 1.0656 $\pm$ 0.1000 \\
1.2809 $\pm$ 0.1000 & 1.5882 $\pm$ 0.1000 & 1.2809 $\pm$ 0.1000 \\
1.0556 $\pm$ 0.1000 & 0.9513 $\pm$ 0.1000 & 1.0556 $\pm$ 0.1000 \\
1.0444 $\pm$ 0.1000 & 1.4675 $\pm$ 0.1000 & 1.0444 $\pm$ 0.1000 \\
1.1543 $\pm$ 0.1000 & 1.0108 $\pm$ 0.1000 & 1.1543 $\pm$ 0.1000 \\
0.9276 $\pm$ 0.1000 & 0.5004 $\pm$ 0.1000 & 0.9276 $\pm$ 0.1000 \\
0.8812 $\pm$ 0.1000 & 1.0645 $\pm$ 0.1000 & 0.8812 $\pm$ 0.1000 \\
0.9702 $\pm$ 0.1000 & 0.7038 $\pm$ 0.1000 & 0.9702 $\pm$ 0.1000 \\
0.9339 $\pm$ 0.1000 & 1.3435 $\pm$ 0.1000 & 0.9339 $\pm$ 0.1000 \\
0.9953 $\pm$ 0.1000 & 0.7225 $\pm$ 0.1000 & 0.9953 $\pm$ 0.1000 \\
0.8223 $\pm$ 0.1000 & 0.6221 $\pm$ 0.1000 & 0.8223 $\pm$ 0.1000 \\
0.9966 $\pm$ 0.1000 & 1.4680 $\pm$ 0.1000 & 0.9966 $\pm$ 0.1000 \\
1.0476 $\pm$ 0.1000 & 1.3121 $\pm$ 0.1000 & 1.0476 $\pm$ 0.1000 \\
1.0147 $\pm$ 0.1000 & 0.6549 $\pm$ 0.1000 & 1.0147 $\pm$ 0.1000 \\
1.0069 $\pm$ 0.1000 & 1.4602 $\pm$ 0.1000 & 1.0069 $\pm$ 0.1000 \\
1.1508 $\pm$ 0.1000 & 1.0762 $\pm$ 0.1000 & 1.1508 $\pm$ 0.1000 \\
-& -& 1.5000 $\pm$ 0.0333 \\
\end{tabular}
\end{ruledtabular}
\end{table}

\bibliographystyle{iopart-num}
\bibliography{bayesian_average}

\end{document}